\documentclass[useAMS,usenatbib,usegraphicx]{mn2e}
\onecolumn{}

\title[KIC 6220497: a Pulsating Eclipsing Binary]{KIC 6220497: A New Algol-type Eclipsing Binary with Multiperiodic Pulsations}

\author[Jae Woo Lee et al.]
       {Jae Woo Lee$^{1,2}$\thanks{E-mail: jwlee@kasi.re.kr}, Kyeongsoo Hong$^{1}$, Seung-Lee Kim$^{1,2}$ and Jae-Rim Koo$^{1}$ \\
        $^1$Korea Astronomy and Space Science Institute, Daejeon 34113, Korea \\
        $^2$Astronomy and Space Science Major, Korea University of Science and Technology, Daejeon 34113, Korea }

\begin{document}

\date{Accepted 2016 ---------. Received 2016 ---------; in original form 2016 }

\pagerange{\pageref{firstpage}--\pageref{lastpage}} \pubyear{2016}

\maketitle

\label{firstpage}

\begin{abstract}
We present both binarity and pulsation of KIC 6220497 from the {\it Kepler} observations. The light curve synthesis shows that 
the eclipsing system is a semi-detached Algol with parameters of $q$ = 0.243$\pm$0.001, $i$ = 77.3$\pm$0.3 deg, and 
$\Delta T$ = 3,372$\pm$58 K, in which the detached primary component fills its Roche lobe by $\sim$87\%. A multiple 
frequency analysis of the eclipse-subtracted light residuals reveals 33 frequencies in the range of 0.75$-$20.22 d$^{-1}$ 
with amplitudes between 0.27 and 4.56 mmag. Among these, four are pulsation frequencies in fundamental ($f_1$, $f_5$) and 
$p$ ($f_2$, $f_7$) modes, and six are orbital frequency ($f_8$, $f_{31}$) and its harmonics ($f_6$, $f_{11}$, $f_{20}$, $f_{24}$), 
which can be attributed to tidally excited modes. For the pulsation frequencies, the pulsation constants of 0.16$-$0.33 d and 
the period ratios of $P_{\rm pul}/P_{\rm orb}$ = 0.042$-$0.089 indicate that the primary component is a $\delta$ Sct pulsating 
star and, thus, KIC 6220497 is an oscillating eclipsing Algol (oEA) star. The dominant pulsation period of 0.1174051$\pm$0.0000004 d 
is significantly longer than that expected from empirical relations that link the pulsation period with the orbital period. 
The surface gravity of $\log g_1$ = 3.78$\pm$0.03 is clearly smaller than those of the other oEA stars with 
similar orbital periods. The pulsation period and the surface gravity of the pulsating primary demonstrate that KIC 6220497 
would be the more evolved EB, compared with normal oEA stars.
\end{abstract}


\section{INTRODUCTION}

$\delta$ Sct stars are main sequence and subgiant stars with spectral types from about A2 to F2 (Rodr\'iguez \& Breger 2001). 
They pulsate in low-order radial/non-radial pressure ($p$) modes with short periods of 0.02$-$0.2 d (Breger 2000). The pulsations are 
driven by the $\kappa$ mechanism acting in the He II partial ionization region. Recently, Balona (2014) showed that nearly all 
$\delta$ Sct stars pulsate in the frequency range of 0$-$5 d$^{-1}$ characteristic of $\gamma$ Dor stars and suggested that 
the low frequencies are directly associated with the presence of $\delta$ Sct pulsations. The $\delta$ Sct and $\gamma$ Dor variables 
share a similar parameter space in the Hertzsprung-Russell (HR) diagram which is partly overlapped, but the latter stars pulsate 
in high-order non-radial gravity ($g$) modes driven by a mechanism known as convective blocking (Guzik et al. 2000) with relatively 
longer periods of 0.4$-$3 d (Handler \& Shobbrook 2002; Henry, Fekel \& Henry 2005). The hybrid stars that contain the $p$ and 
$g$ modes simultaneously are of particular interest, since they probe both the envelope and the deep interior near the core region 
of the pulsators (Kurtz et al. 2015). 

Over 90 pulsating stars have been known as the $\delta$ Sct, $\gamma$ Dor, and hybrid stars in eclipsing binaries (EBs). 
Approximately 73 of them are the so-called oEA (oscillating eclipsing Algol) stars; the mass-accreting components of classical 
semi-detached Algols that lie inside the instability strip and show $\delta$ Sct-like oscillations (Mkrtichian et al. 2004). 
The $\delta$ Sct stars in EBs exhibit almost the same pulsating characteristics as single $\delta$ Sct pulsators, but 
their evolutionary processes are entirely different from each other by tidal interaction and mass transfer between components. 
For the pulsating EBs, a possible relation between the binary orbital periods ($P_{\rm orb}$) and the dominant pulsation periods 
($P_{\rm pul}$) was firstly given by Soydugan et al. (2006) and updated by Liakos et al. (2012): 
$P_{\rm pul}$ = 0.020$P_{\rm orb}$ $-$ 0.005 for 20 EBs and $\log P_{\rm pul}$ = 0.58$P_{\rm orb}$ $-$ 1.53 for 70 EBs in 
the same order. Later, the $P_{\rm pul}-P_{\rm orb}$ relation was theoretically established by Zhang, Luo \& Fu (2013), in which 
the pulsation period could be described as a function of the orbital period, the pulsation constant, the mass ratio, and 
the filling factor. According to those studies, the eclipsing $\delta$ Sct stars pulsate with shorter pulsational periods than 
single $\delta$ Sct pulsators. The pulsational period seems to depend on the gravitational force exerted by a companion onto 
the pulsating star. This indicates that the companion star may influence the pulsation frequencies and modes of the pulsating component. 
The pulsating stars in EBs offer a unique opportunity to study the effects of tidal forces and companions on the pulsations 
(Hambleton et al. 2013), as well as to measure directly the masses and radii of the pulsators from binary modelling.

The EBs with pulsating stars are the Rosetta stone for the study of stellar structure and evolution through asteroseismology
and binary properties. Such examples are KIC 10661783 (Southworth et al. 2011; Lehmann et al. 2013), 
KIC 4544587 (Hambleton et al. 2013), and KIC 3858884 (Maceroni et al. 2014). Because space missions such as {\it Kepler} and 
{\it CoRot} provide ultra-precise photometric data, they allow the detection of many pulsation frequencies with amplitudes down 
to the micromagnitude level. In order to look for pulsating components in EBs and to understand their physical properties, 
we choose the {\it Kepler} target KIC 6220497 (R.A.$_{2000}$ = 19$^{\rm h}$44$^{\rm m}$39$\fs547$; decl.$_{2000}$ = +41$^{\circ}$33${\rm '}$21$\farcs$17; 
$K_{\rm p}$ = $+$14.749; $g$=$+$14.935; $g-r$=$+$0.247), which was announced to be probably a pulsating EB with an orbital period 
of 1.323 d by Gaulme \& Guzik (2014). This paper is the third contribution in a series assessing the detections and properties of 
pulsating stars in the {\it Kepler} EBs (Lee et al. 2014, 2016). Here, we present the binary system as an oEA star showing 
multiperiodic $\delta$ Sct pulsations, based on light-curve synthesis and frequency analysis for the light residuals from 
the binary model.

\section{{\it KEPLER} PHOTOMETRY AND LIGHT-CURVE SYNTHESIS}

KIC 6220497 was observed during Quarters 14 and 15 in the long cadence mode, which has a sampling time of 29.4 min. We used
the data in the {\it Kepler} EB catalogue\footnote{http://keplerebs.villanova.edu/} detrended and normalised from the raw SAP 
(Simple Aperture Photometry) time series (Pr\v sa et al. 2011; Slawson et al. 2011). The contamination level of the measurements 
is estimated to be 0.046. This value suggests that the {\it Kepler} target suffers minimally from third light, if any. Figure 1 
depicts the {\it Kepler} light curve for KIC 6220497, where there are no significant trends in the eclipse depths and 
the light maxima (Max I and Max II) present equal light levels. As shown in the phase-folded light curve, the binary star 
displays a considerable variation outside eclipses due to tidal distortions. The depth difference between the primary and 
secondary eclipses indicates a large temperature difference between the component stars. 

For the light-curve synthesis of KIC 6220497, we used the 2007 version of the Wilson-Devinney synthesis code 
(Wilson \& Devinney 1971, van Hamme \& Wilson 2007; hereafter W-D). The {\it Kepler} data of this system were analysed in 
a manner almost identical to that for the pulsating EBs V404 Lyr (KIC 3228863; Lee et al. 2014) and KIC 4739791 
(Lee et al. 2016). The effective temperature ($T_1$) of the brighter, and presumably more massive, star was initialised to be 
7,254 K from the {\it Kepler} Input Catalogue (KIC; Kepler Mission Team 2009). The logarithmic bolometric ($X_{1,2}$) 
and monochromatic ($x_{1,2}$) limb-darkening coefficients were interpolated from the values of van Hamme (1993). 
The gravity-darkening exponents were fixed at standard values of $g_1$=1.0 and $g_2$=0.32, while the bolometric albedos at 
$A_1$=1.0 and $A_2$=0.5, as surmised from the components' temperatures. Adjustable parameters were the orbital ephemeris 
($T_0$ and $P$), the mass ratio ($q$), the orbital inclination ($i$), the effective temperatures ($T_{1,2}$) and 
the dimensionless surface potentials ($\Omega_{1,2}$) of the components, and the monochromatic luminosity ($L_{1}$).
This synthesis was repeated until the correction of each adjustable parameter became smaller than its standard deviation using 
the differential correction (DC) programme of the W-D code. In this paper, the subscripts 1 and 2 refer to the primary and 
secondary components being eclipsed at Min I and Min II, respectively. 

The mass ratio ($q$ = $M_2$/$M_1$) is a very important parameter in the light-curve synthesis. However, there exist neither 
photometric solutions nor spectroscopic orbits for KIC 6220497. Because the binary star is a faint object with a short orbital 
period, 8$-$10 m class telescopes are needed to measure its radial velocities. However, it is very difficult to obtain 
such observation times. Thus, we conducted an extensive $q$-search procedure, meaning that we calculated a series of 
models with the mass ratios in step of 0.01 between 0.1 and 1.0. For each assumed mass ratio, the DC programme was applied 
for various modes but showed acceptable photometric solutions only for semi-detached mode 5 in which the secondary component 
fills its inner Roche lobe. As displayed in Figure 2, the $q$ searches indicate that the minimum value of the weighted sum of 
the squared residuals ($\sum{W(O-C)^2}$; hereafter $\sum$) is around $q$ = 0.24. In the subsequent calculations, this $q$ value 
was treated as an adjustable parameter. The result is listed in the second and third columns of Table 1 and appears as 
a solid curve in the right panel of Figure 1. The light residuals from Model 1 are plotted as the grey circles in the lower panel 
of Figure 3, wherein it can be seen that the model light curves describe the {\it Kepler} data satisfactorily. 

The photometric solutions for Model 1 could be affected by the multiperiodic pulsations of the primary component, which will be 
discussed in the following section. We removed the pulsation signatures from the original {\it Kepler} data, leaving 
behind the light variations due to binarity effects. Then, the new light curve was solved by using Model 1 as initial values. 
The final result is illustrated in Figure 3, and is given as Model 2 in the fourth and fifth columns of Table 1. As listed in 
the table, the binary parameters for Model 2 are in good agreement with those for Model 1. We can see that the photometric 
solutions of KIC 6220497 are immune from the light variation due to the pulsations. Our light-curve synthesis represents 
KIC 6220497 as a semi-detached eclipsing system in which the primary component fills its limiting lobe by 
$\Omega_1$/$\Omega_{\rm in}$ = 87\%, where $\Omega_{\rm in}$ is the potential for the inner critical surface. The Roche-geometry 
configuration of the system permits some mass transfer from the lobe-filling secondary to the detached primary component. 
In all the procedures that have been described, we included the orbital eccentricity ($e$) as a free parameter but found that 
the parameter remained indistinguishable from zero within its error. This indicates that KIC 6220497 is in a circular orbit, 
as expected for semi-detached classical Algols. 

It is known that the stand errors of the adjustable parameters taken from the W-D code are unrealistically small because 
of the strong correlations between relatively many parameters and partly the non-normal distribution of measurement errors 
(Maceroni \& Rucinski 1997). In order to obtain more reliable errors for the Model 1 and Model 2 parameters, we followed 
the procedure described by Koo et al. (2014). First of all, we divided the observed (or prewhitened) {\it Kepler} data of 
KIC 6220497 into 138 segments at the interval of an orbital period and separately analysed them with the W-D code. Then, 
we computed the standard deviations of each parameter from the 138 different values. The parameter errors presented in Table 1 
are the 1$\sigma$-values adopted from this procedure. 

The temperatures of the primary component in Table 1 correspond to a normal main-sequence star with a spectral type of A9. Because 
the temperature errors are certainly underestimated, it was assumed that the temperature of each component had an error of 200 K. 
Using the correlations between spectral type and stellar mass (Harmanec 1988), we estimated the primary's mass to be 
$M_1$=1.60$\pm$0.08$M_\odot$. The absolute dimensions for KIC 6220497 can be computed from our photometric solutions and $M_1$. 
These are given in the bottom of Table 1, where the radii are the mean volume radii calculated from the tables of Mochnacki (1984). 
The luminosities and the bolometric magnitudes were computed by adopting $T_{\rm eff}$$_\odot$=5,780 K and $M_{\rm bol}$$_\odot$=+4.73 
for solar values.

\section{LIGHT RESIDUALS AND PULSATIONAL CHARACTERISTICS}

In order to obtain more reliable frequencies, the observed {\it Kepler} data were split into 138 subsets as before and 
modeled individually with the W-D code through adjusting only the ephemeris epoch ($T_0$) in the Model 1 of Table 1. The light 
residuals from the analyses are displayed in Figure 4 as magnitudes versus BJDs, wherein the lower panel presents a short section 
of the residuals. Light variations with a total peak-to-peak amplitude of $\sim$10 mmag are clearly seen in the residuals. 
In order to examine the oscillating features, we applied a multiple frequency analysis to the whole data sets at once. 
The PERIOD04 programme by Lenz \& Breger (2005) was performed on the frequency range from 0 to the Nyquist limit of 24.47 d$^{-1}$. 
Because the primary component lies within the $\delta$ Sct instability strip of the HR diagram, it would be a candidate for 
such pulsations. During primary eclipses, the secondary component partially blocks the lights of the pulsating primary star, which 
can have an effect on the observed frequencies and amplitudes. In contrast, the light contribution from the secondary is only 2.2 \% 
to the total luminosity of the binary system, so the pulsations are almost unaffected during the secondary eclipses. Thus, we made 
use of only the light residuals having orbital phases between 0.12 and 0.88 after eliminating the data of the primary eclipses. 

The amplitude spectra for KIC 6220497 are shown in the top panel of Figure 5. After the successive prewhitening of each frequency 
peak, we detected 33 frequencies with the signal to noise amplitude (S/N) ratios larger than 4.0 (Breger et al. 1993). At each step 
of this procedure, a multiperiodic least-squares fit to the light residuals was carried out using the fitting formula of 
$Z$ = $Z_0$ + $\Sigma _{i}$ $A_i \sin$(2$\pi f_i t + \phi _i$). Here, $Z$ and $Z_0$ denote the calculated magnitude and zero point, 
respectively, $A_i$ and $\phi _i$ are the amplitude and phase of the $i$th frequency, respectively, and $t$ is the time of 
each measurement. The amplitude spectra after prewhitening the first seven frequencies and then all 33 frequencies are presented 
in the middle and bottom panels of Figure 5, respectively. The results are presented in Table 2, where the uncertainties were 
calculated according to Kallinger, Reegen \& Weiss (2008). The synthetic curve obtained from the 33-frequency fit is displayed in 
the lower panel of Figure 4. Some additional peaks still exist in the bottom panel of Figure 5, but their S/N ratios are smaller 
than the empirical threshold of 4.0. 

As in the case of KIC 4739791 (Lee et al. 2016), to see if the main frequencies detected in this paper are real and stable during 
the observing run of $\sim$ 200 d, we re-analysed the light residuals at intervals of about 50 d and examined the frequency 
variations with time. The four subsets resulted in slight differences from each other and approximate 20 frequencies were 
detected at each subset with the same criterion of S/N$>$4.0. Figure 6 showed the stability of the 14 frequencies. Among these, 
the four frequencies ($f_9$, $f_{10}$, $f_{14}$, $f_{21}$) varied significantly and the three frequencies decreased ($f_3$, $f_{16}$) 
or increased ($f_{17}$) monotonically. Within the frequency resolution of 0.008 d$^{-1}$ (Loumos \& Deeming 1978), we checked 
the frequencies for possible harmonic and combination terms. The result is listed in the last column of Table 2, where the six 
($f_6$, $f_8$, $f_{11}$, $f_{20}$, $f_{24}$, $f_{31}$) frequencies are the orbital frequency ($f_{\rm orb}$ = 0.75576 d$^{-1}$) 
and its harmonics.

\section{DISCUSSION AND CONCLUSIONS}

In this paper, we studied the physical properties of KIC 6220497 from detailed analyses of the {\it Kepler} time-series data 
obtained during Quarters 14 and 15. The ${\it Kepler}$ light curve was satisfactorily modelled for two cases: including and 
removing pulsations. The binary parameters between them are consistent with each other, which indicates that the photometric 
analysis of KIC 6220497 is almost unaffected by pulsations. The light-curve synthesis presented in this paper demonstrates that 
the system is a semi-detached EB with parameters of $q$ = 0.243$\pm$0.001, $i$ = 77.3$\pm$0.3 deg, and 
$\Delta T$ = 3,372$\pm$58 K. The detached primary component is about 1.6 times larger than the lobe-filling secondary and 
fills its inner critical lobe by about 87\%, which is one of the largest filling factors for pulsating EBs. In Figure 7, 
a comparison of the KIC 6220497 parameters with the mass-radius, mass-luminosity, and HR diagrams shows that the primary star 
lies in the main-sequence band, while the secondary is highly evolved and its radius and luminosity are more than four times 
oversized and about 26 times overluminous, respectively, compared with dwarf stars of the same mass. In these diagrams, 
the locations of both components conform to the general pattern of semi-detached Algols (\. Ibano\v{g}lu et al. 2006). 

For the detection of the pulsation frequencies in KIC 6220497, we removed the binarity effects from the observed {\it Kepler} data 
and performed a frequency analysis in the entire light residuals, excluding the data around the primary minima. As a consequence, 
we detected the 33 frequencies including a dominant oscillation found at $f_{1}$ = 8.51752$\pm$0.00003 d$^{-1}$, corresponding to 
0.1174051$\pm$0.0000004 d. Among these, the four ($f_1$, $f_2$, $f_5$, $f_7$) frequencies in the $p$-mode region were highly stable 
during the observing runs of about 200 d. On the contrary, the other frequencies varied with time and/or may be orbital harmonics 
and combination terms. The frequency ratio of $f_1$ and $f_5$ is 0.984, very close to 1.0, and these frequencies could not be 
interpreted by two radial modes of $\delta$ Sct stars (Breger 1979). We checked the ratios of the other frequencies and found 
one candidate to be identified as radial modes, i.e. the ratio of 0.848 between $f_2$ and $f_7$ is nearly the same with 
the period ratio of 0.845 for the second (2H) and third (3H) overtone radial modes of the $\delta$ Sct stars. It means that 
the longer period $f_7$ could be classified as the second overtone radial mode and the shorter period $f_2$ as the third overtone 
radial mode. Applying the physical parameters of the primary component in Table 1 to the equation of 
$\log Q_i = -\log f_i + 0.5 \log g + 0.1M_{\rm bol} + \log T_{\rm eff} - 6.456$ (Breger 2000), we obtained the pulsation constants 
for the four frequencies ($f_{1,2,5,7}$) to be $Q_1$ = 0.033 d, $Q_2$ = 0.016 d, $Q_5$ = 0.033 d, and $Q_7$ = 0.019 d. We compared 
these $Q$ values with the theoretical models with 1.5 $M_\odot$ given by Fitch (1981). The $f_1$ and $f_5$ frequencies could be 
identified as fundamental ($n = 0$) modes and the $f_2$ and $f_7$ as the third ($n = 3$) and second ($n = 2$) overtone $p$-modes, 
respectively. The ratios, $P_{\rm pul}/P_{\rm orb}$ = 0.042$\sim$0.089, of the  pulsational to orbital periods are within 
the upper limit of $P_{\rm pul}/P_{\rm orb}$ = 0.09$\pm$0.02 for $\delta$ Sct stars in EBs that could be used to distinguish 
approximately whether a binary component pulsates in the $p$-mode (Zhang, Luo \& Fu 2013). These results indicate that 
the primary star with a spectral type of A9V can be classified as a $\delta$ Sct variable. Because the system is in 
a semi-detached configuration, these results imply that KIC 6220497 would be an oEA star and a mass transfer from the evolved 
lobe-filling secondary to the detached primary component could be responsible for the $\delta$ Sct-type pulsations detected in 
this paper. 

The oEA stars have pulsation features similar to classical $\delta$ Sct stars. However, their pulsations may be influenced 
by the tidal interaction and mass transfer between the components, as well as gravitational force from companions. Some pulsations
in KIC 6220497 can be excited by the tidal forces of the secondary companion. Tidally excited modes occur when the orbital 
frequency is close to a stellar eigenfrequency in a binary star with an eccentric orbit. The signature of the pulsation modes 
is the frequencies at integer multiples of the orbital frequency (Welsh et al. 2011; Thompson et al. 2012; Hambleton et al. 2013). 
We detected four ($f_6$, $f_{11}$, $f_{20}$, $f_{24}$) frequencies that are the harmonics of the orbital frequency, which could be 
partly affected by imperfect removal of the eclipses from the light curve. Although the orbit of the binary system is circular, 
these frequencies could result from tidally induced pulsations (Reyniers \& Smeyers 2003a,b; Southworth et al. 2011). 
In the upper panel of Figure 8, we plot the dominant pulsation period versus the orbital period for 74 oEA stars, including 
KIC 6220497. The data are taken from the compilations of Zhang, Luo \& Fu (2013; 67 oEA stars) and from more recent literature 
(Yang, Wei \& Li 2014 for FR Ori; Zhang et al. 2014 for OO Dra; Zhang, Luo \& Wang 2015 for EW Boo; Zhang et al. 2015 for V392 Ori; 
Lee et al. 2016 for KIC 4739791; Soydugan et al. 2016 for XZ Aql). In the panel, we can see that the pulsation period of KIC 6220497 
deviated from the general trend of the oEA stars and also the empirical relation between $P_{\rm pul}$ and $P_{\rm orb}$ of 
Zhang, Luo \& Fu (2013). The pulsation periods in EBs increase with decreasing gravitational pull exerted by the secondary 
companion to the pulsating primary component (Soydugan et al. 2006). For KIC 6220497, the gravitational pull applied to per gram 
of the matter on the surface of the pulsating component by the lobe-filling secondary was calculated to be $\log F$ = 2.90 in 
cgs units, which is about 3.4 times smaller than the value of 9.89 (again, in cgs units) taken from the relation of 
$\log P_{\rm pul} = -0.61 \log F +$ 5.1 calibrated by Soydugan et al. (2006). However, as shown in the lower panels of Figure 8 
from 34 oEA stars currently known, the gravitational force $\log F$ for KIC 6220497 matches well with those for the other oEA stars 
with similar orbital periods, while the surface gravity $\log g_1$ = 3.78 for the primary component is clearly smaller. 
The period-gravity relation is similar to that for radially pulsating stars suggested by Fernie (1995); as the surface gravity 
decreases, its pulsation period increases. Further, it is known that the more evolved the star, the slower the pulsations 
(cf. Liakos et al. 2012). The pulsation period and the surface gravity of KIC 6220497 indicate that the system might be a more 
evolved EB than the other oEA stars. We think that the pulsation periods strongly depend on the surface gravities of the pulsating 
components and the evolutionary status of the binary stars.

When the double-lined radial velocities and the multiband light curves are made (e.g., Hong et al. 2015, Koo et al. 2016), 
they will help to understand the absolute parameters, identification of pulsation modes, and evolutionary status of KIC 6220497 
better than now. These offer us important information for asteroseismology and the study of stellar interior and evolution.

\section*{Acknowledgments}
We appreciate the careful reading and valuable comments of the anonymous referee. This paper includes data collected by 
the {\it Kepler} mission. {\it Kepler} was selected as the 10th mission of the Discovery Program. Funding for the {\it Kepler} 
mission is provided by the NASA Science Mission directorate. We have used the Simbad database maintained at CDS, Strasbourg, 
France. This work was supported by the KASI (Korea Astronomy and Space Science Institute) grant 2016-1-832-01.

\clearpage
\begin{figure}
\includegraphics[scale=0.85]{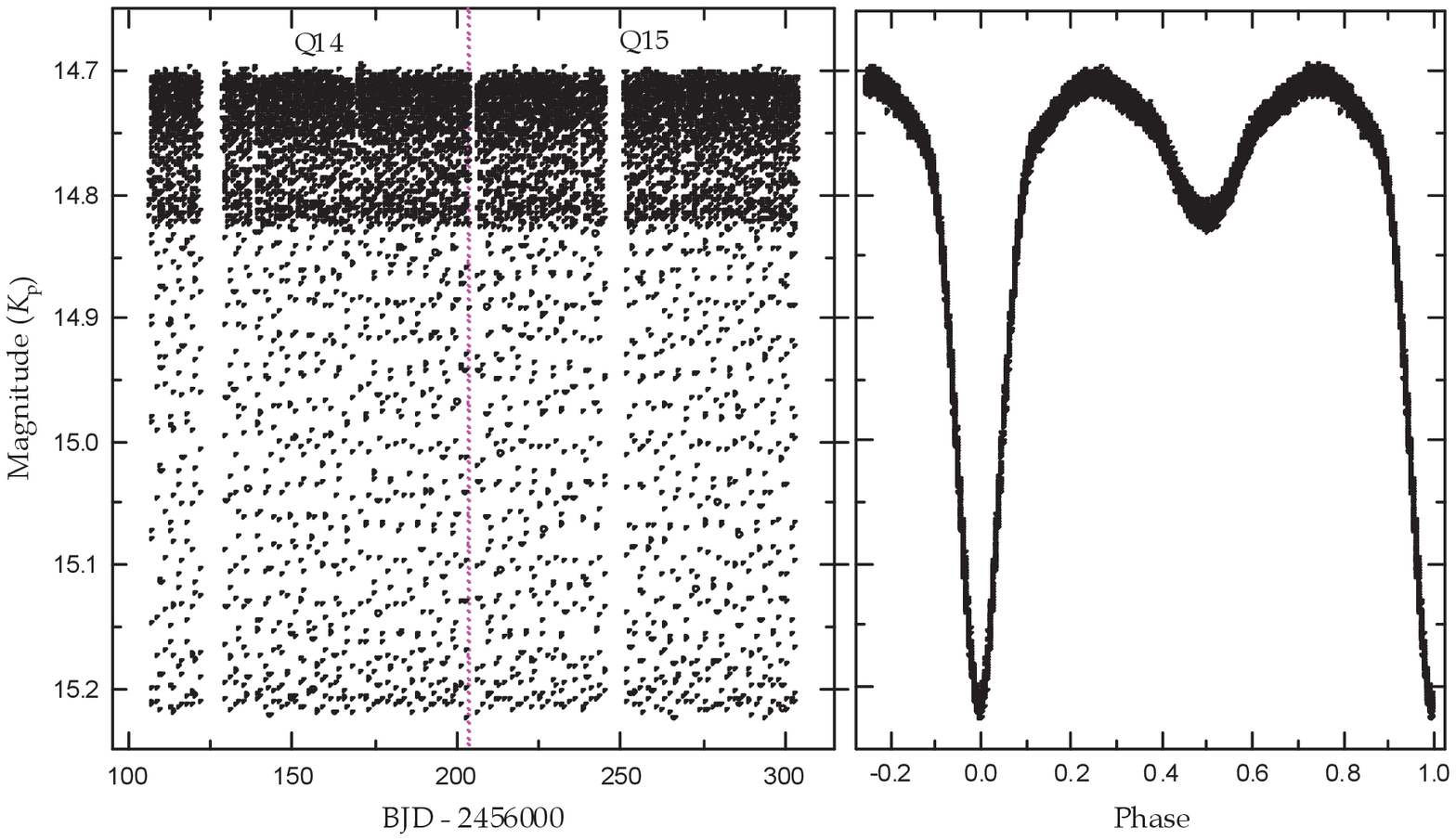}
\caption{Observed {\it Kepler} light curve of KIC 6220479 distributed in BJD (left panel) and orbital phase (right panel). 
In the left panel, the vertical dashed line represents the ending time of Quarter 14. }
\label{Fig1}
\end{figure}

\begin{figure}
\includegraphics[]{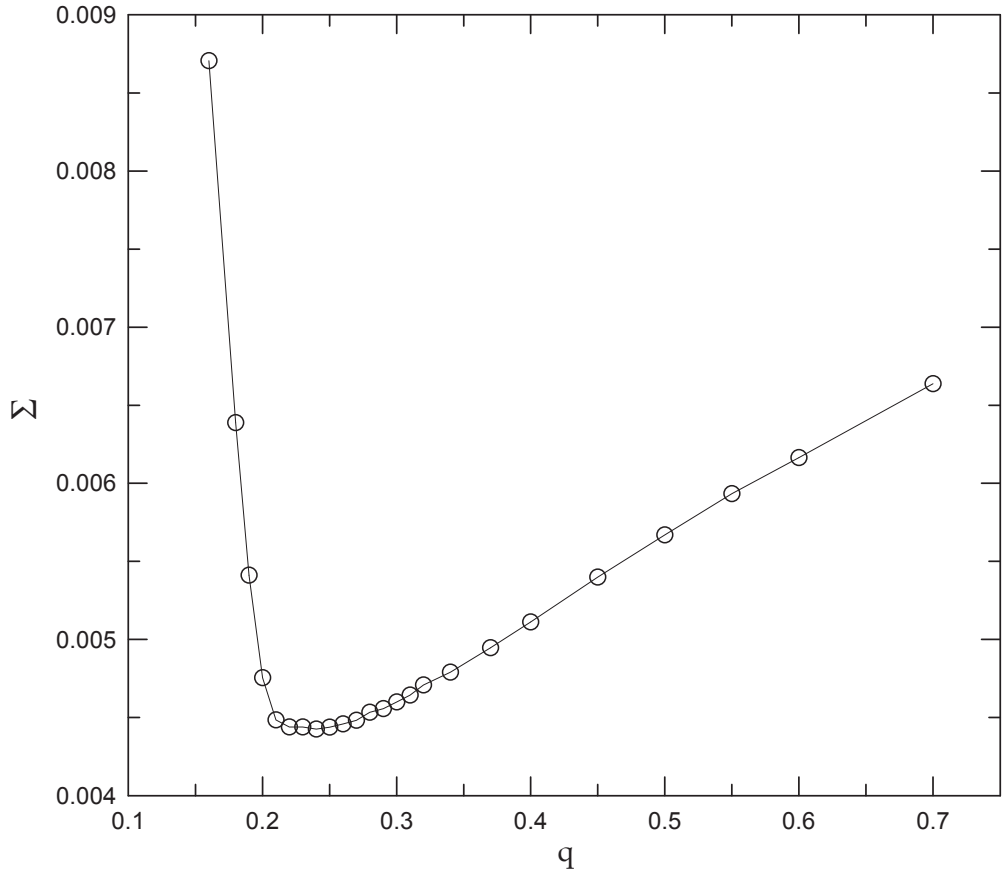}
\caption{Behaviour of $\sum$ (the sum of the residuals squared) of KIC 6220497 as a function of mass ratio $q$, showing 
a minimum value at $q$=0.24. The circles represent the $q$-search results for each assumed mass ratio. }
\label{Fig2}
\end{figure}

\begin{figure}
\includegraphics[]{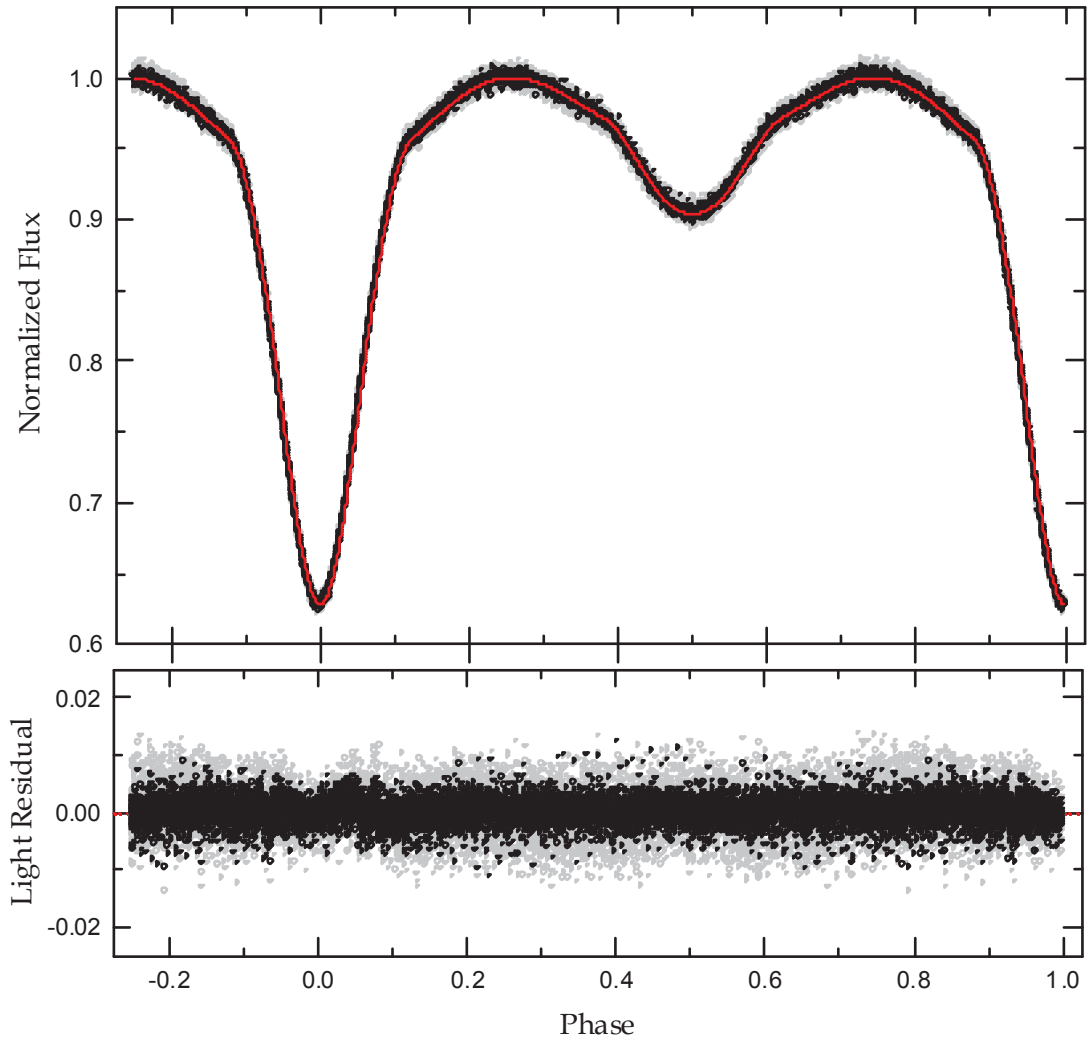}
\caption{Binary light curve (black circle) after subtracting the pulsation signatures from the observed {\it Kepler} data 
(grey circle). In the upper panel, the blue dashed and red solid curves are computed with the Model 1 and Model 2 parameters 
of Table 1, respectively. Because the two models are in substantial agreement with each other, the fitted Model 1 cannot be seen 
individually. The lower panel represents the light residuals from both datasets: observed and prewhitened light curves. }
\label{Fig3}
\end{figure}

\begin{figure}
\includegraphics[]{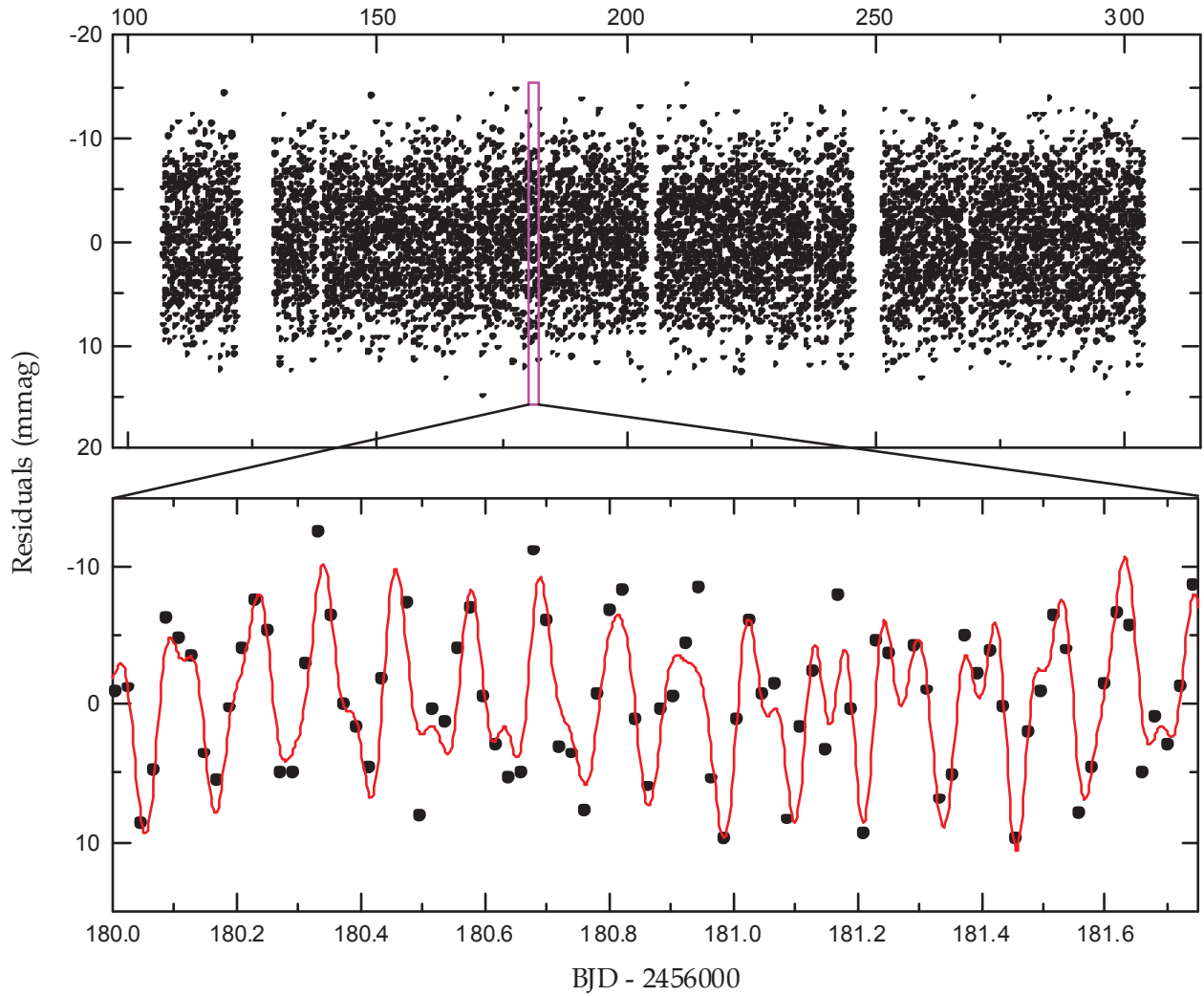}
\caption{Light residuals after subtracting the binarity effects from each of the 138 light curves in the original {\it Kepler} data. 
The lower panel presents a short section of the residuals marked using the inset box of the upper panel. The synthetic curve was 
computed from the 33-frequency fit to the data between orbital phases 0.12 and 0.88. }
\label{Fig4}
\end{figure}

\begin{figure}
\includegraphics[]{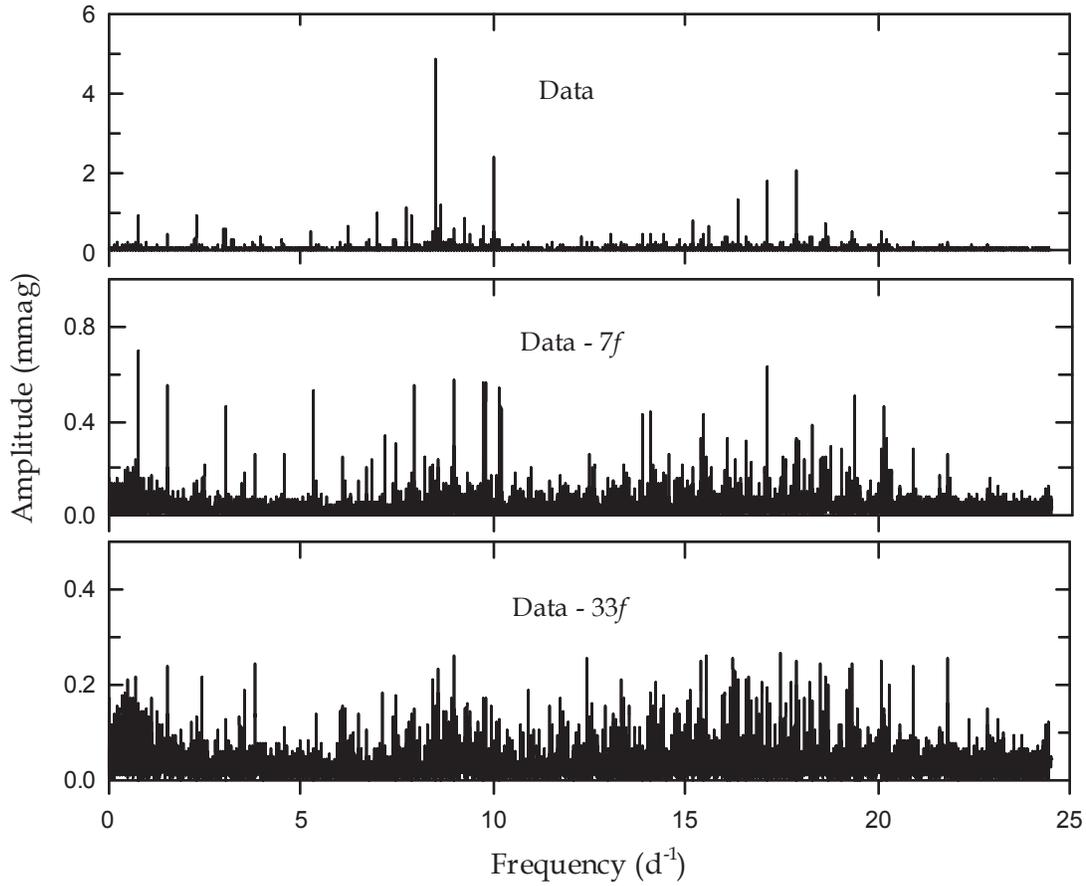}
\caption{Amplitude spectra before (top panel) and after prewhitening the first 7 frequencies (middle panel) and all 33 frequencies 
(bottom panel) from the PERIOD04 programme. The frequency analysis was applied to the entire {\it Kepler} data except for 
the times of the primary eclipses. }
\label{Fig5}
\end{figure}

\begin{figure}
\includegraphics[]{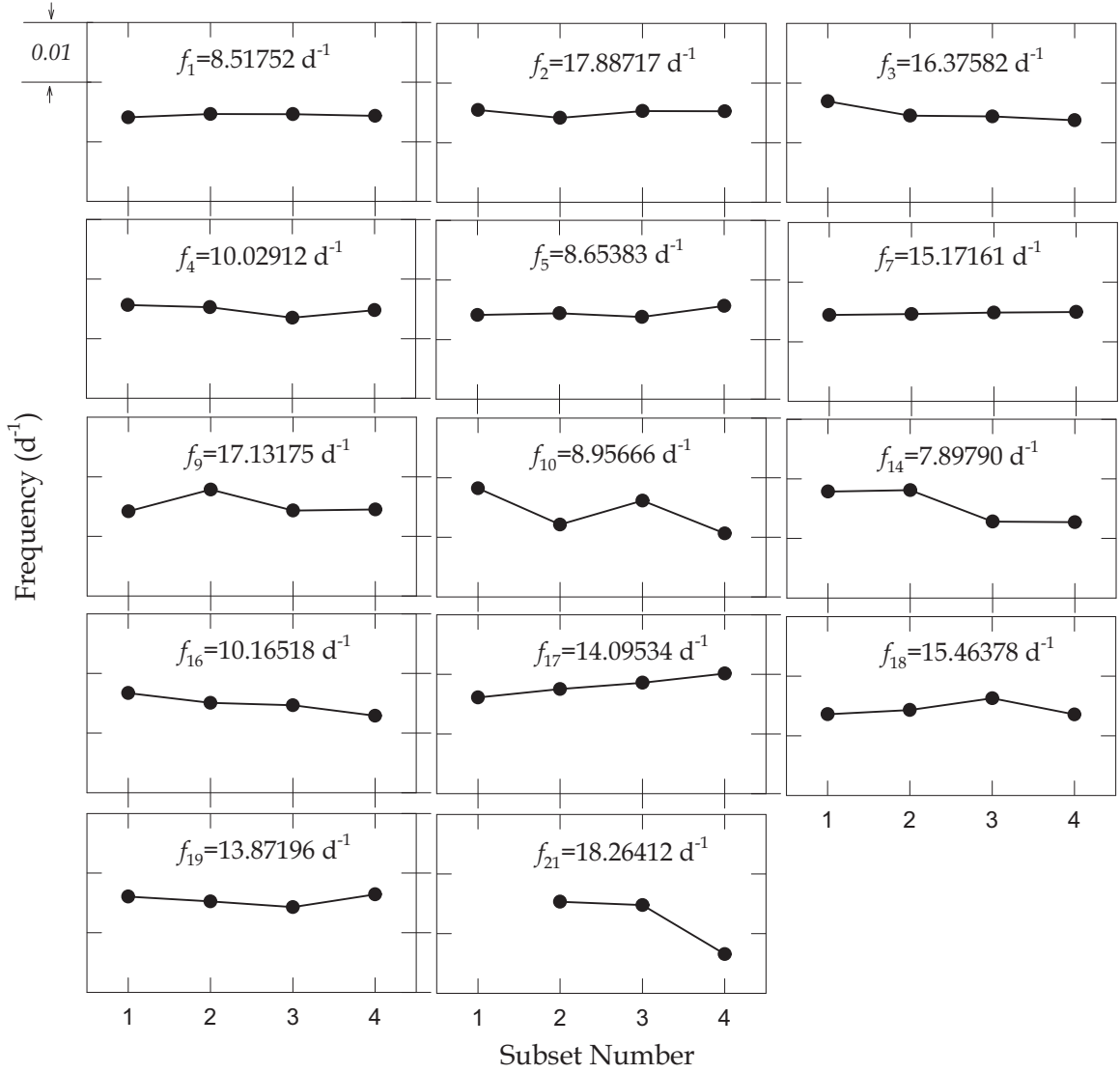}
\caption{Variability of the main frequencies detected in the four subsets at intervals of approximately 50 d. In all panels, 
the y-axes are scaled to 0.03 d$^{-1}$, and the tick intervals are 0.01 d$^{-1}$. }
\label{Fig6}
\end{figure}

\begin{figure}
\includegraphics[]{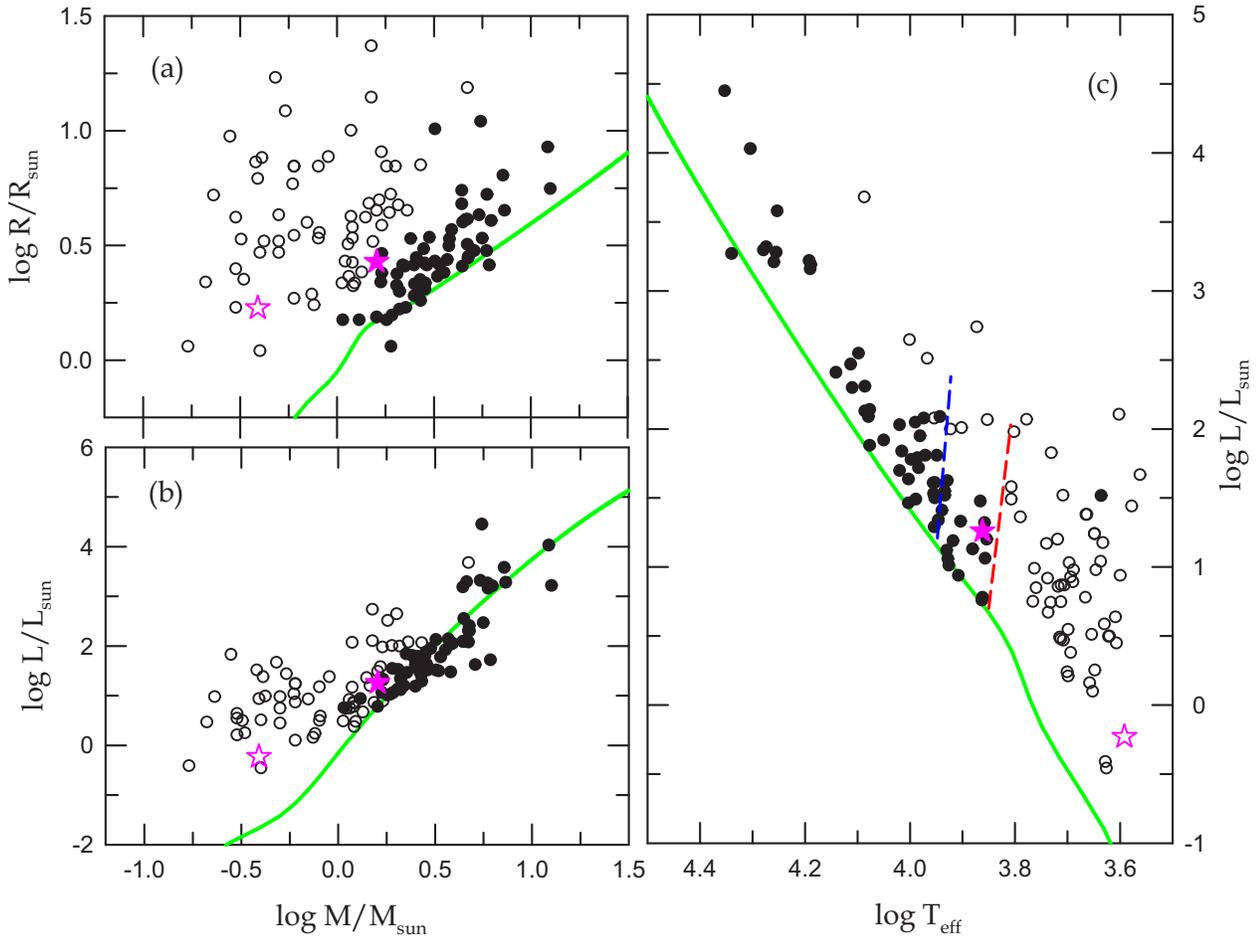}
 \caption{Plots of (a) mass-radius, (b) mass-luminosity, and (c) HR diagrams for semi-detached Algols (\. Ibano\v{g}lu et al. 2006). 
The filled and open circles represent the primary and secondary components, respectively. The filled and open star symbols denote 
the locations of the primary and secondary components of KIC 6220497, respectively, and the solid lines represent the ZAMS stars 
calculated as having a solar metallicity of $Z$=0.02 found in Tout et al. (1996). In panel (c), the dashed blue and red lines 
represent the theoretical edges of the $\delta$ Sct instability strip, and the pulsating primary star of KIC 6220497 resides within 
the region. }
\label{Fig7}
\end{figure}

\begin{figure}
\includegraphics[]{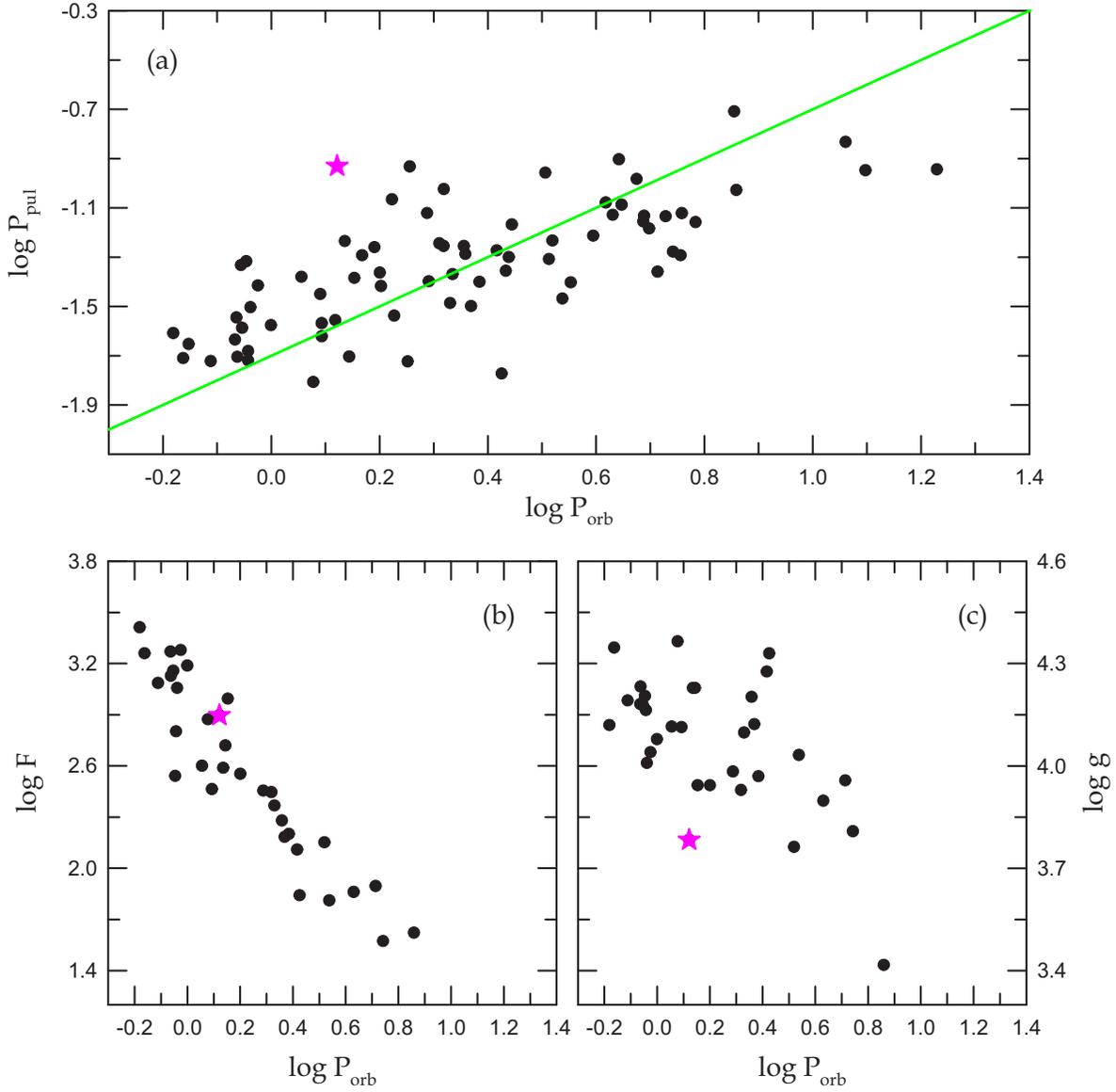}
\caption{Plots of (a) pulsation periods $\log P_{\rm pul}$, (b) the gravitational forces $\log F$, and (c) surface gravities $\log g$ 
against the orbital periods $P_{\rm orb}$ for oEA stars. The star symbols denote the parameters of KIC 6220497, and the solid line 
in the upper panel represents the relation of $\log P_{\rm pul} = \log P_{\rm orb} -$ 1.70 given by Zhang, Luo \& Fu (2003). 
See text for details. }
\label{Fig8}
\end{figure}

\clearpage
\begin{table}
\caption{Binary Parameters of KIC 6220497. }
\begin{tabular}{lccccc}
\hline
Parameter                                & \multicolumn{2}{c}{Model 1$\rm ^a$}         && \multicolumn{2}{c}{Model 2$\rm ^b$}         \\ [1.0mm] \cline{2-3} \cline{5-6} \\[-2.0ex]
                                         & Primary           & Secondary               && Primary           & Secondary               \\                                                                                         
\hline
$T_0$ (BJD)                              & \multicolumn{2}{c}{2,456,204.39612(5)}      && \multicolumn{2}{c}{2,456,204.39619(2)}      \\
$P$ (day)                                & \multicolumn{2}{c}{1.323167(1)}             && \multicolumn{2}{c}{1.3231670(6)}            \\
$q$                                      & \multicolumn{2}{c}{0.242(3)}                && \multicolumn{2}{c}{0.243(1)}                \\
$i$ (deg)                                & \multicolumn{2}{c}{77.3(1.6)}               && \multicolumn{2}{c}{77.3(3)}                 \\
$T$ (K)                                  & 7,263(64)         & 3,911(20)               && 7,279(54)         & 3,907(22)               \\
$\Omega$                                 & 2.688(13)         & 2.335                   && 2.689(7)          & 2.336                   \\
$A$                                      & 1.0               & 0.5                     && 1.0               & 0.5                     \\
$g$                                      & 1.0               & 0.32                    && 1.0               & 0.32                    \\
$X$, $Y$                                 & 0.640, 0.259      & 0.621, 0.150            && 0.640, 0.259      & 0.621, 0.150            \\
$x$, $y$                                 & 0.598, 0.260      & 0.751, 0.026            && 0.598, 0.258      & 0.750, 0.026            \\
$L$/($L_{1}$+$L_{2}$)                    & 0.978(2)          & 0.022                   && 0.978(2)          & 0.022                   \\
$r$ (pole)                               & 0.4060(15)        & 0.2460(9)               && 0.4058(11)        & 0.2462(6)               \\
$r$ (point)                              & 0.4479(28)        & 0.3590(12)              && 0.4477(18)        & 0.3593(8)               \\
$r$ (side)                               & 0.4251(18)        & 0.2560(9)               && 0.4249(14)        & 0.2562(6)               \\
$r$ (back)                               & 0.4356(22)        & 0.2885(9)               && 0.4355(15)        & 0.2888(6)               \\
$r$ (volume)$\rm ^c$                     & 0.4224(18)        & 0.2644(9)               && 0.4223(13)        & 0.2646(6)               \\ 
$\sum W(O-C)^2$                          & \multicolumn{2}{c}{0.0044}                  && \multicolumn{2}{c}{0.0023}                  \\ [1.0mm]
\multicolumn{6}{l}{Absolute parameters:}                                                                                              \\            
$M$ ($M_\odot$)                          & 1.60(8)           &  0.39(2)                && 1.60(8)           &  0.39(2)                \\
$R$ ($R_\odot$)                          & 2.69(6)           &  1.68(4)                && 2.69(6)           &  1.69(4)                \\
$\log$ $g$ (cgs)                         & 3.78(3)           &  3.57(3)                && 3.78(3)           &  3.57(3)                \\
$L$ ($L_\odot$)                          & 18(2)             &  0.6(1)                 && 18(2)             &  0.6(1)                 \\
$M_{\rm bol}$ (mag)                      & 1.6(1)            &  5.3(2)                 && 1.6(1)            &  5.3(2)                 \\
\hline
\multicolumn{6}{l}{$^a$ Result from the observed {\it Kepler} light curve.} \\
\multicolumn{6}{l}{$^b$ Result from the prewhitened light curve removing the pulsations.} \\
\multicolumn{6}{l}{$^c$ Mean volume radius.} 
\end{tabular}
\end{table}

\begin{table}
\caption{Multiple Frequency Analysis of KIC 6220497$\rm ^a$. }
\begin{tabular}{lrcccc}
\hline
             & Frequency              & Amplitude           & Phase           & S/N$\rm ^b$            & Remark                                \\
             & (day$^{-1}$)           & (mmag)              & (rad)           &                        &                                       \\
\hline
$f_{1}$      &  8.51752$\pm$0.00003   & 4.56$\pm$0.15       & 6.24$\pm$0.10   & 52.39                  &                                       \\ 
$f_{2}$      & 17.88717$\pm$0.00006   & 2.19$\pm$0.13       & 0.77$\pm$0.18   & 28.07                  &                                       \\ 
$f_{3}$      & 16.37582$\pm$0.00009   & 1.56$\pm$0.13       & 4.78$\pm$0.25   & 20.25                  & $f_2-2f_{\rm orb}$                    \\ 
$f_{4}$      & 10.02912$\pm$0.00010   & 1.41$\pm$0.14       & 2.51$\pm$0.30   & 16.79                  & $f_1+2f_{\rm orb}$                    \\ 
$f_{5}$      &  8.65383$\pm$0.00014   & 1.11$\pm$0.15       & 2.04$\pm$0.39   & 12.75                  &                                       \\ 
$f_{6}$      &  2.26728$\pm$0.00015   & 0.62$\pm$0.09       & 5.71$\pm$0.43   & 11.59                  & $3f_{\rm orb}$                        \\ 
$f_{7}$      & 15.17161$\pm$0.00014   & 0.81$\pm$0.11       & 5.44$\pm$0.41   & 12.16                  &                                       \\ 
$f_{8}$      &  0.75567$\pm$0.00014   & 0.78$\pm$0.11       & 3.52$\pm$0.40   & 12.60                  & $f_{\rm orb}$                         \\ 
$f_{9}$      & 17.13175$\pm$0.00017   & 0.81$\pm$0.13       & 2.53$\pm$0.48   & 10.34                  & $f_2-f_{\rm orb}$                     \\ 
$f_{10}$     &  8.95666$\pm$0.00025   & 0.60$\pm$0.15       & 4.52$\pm$0.72   &  6.96                  &                                       \\ 
$f_{11}$     &  5.29073$\pm$0.00014   & 0.58$\pm$0.08       & 1.02$\pm$0.40   & 12.67                  & $7f_{\rm orb}$                        \\ 
$f_{12}$     &  9.73721$\pm$0.00026   & 0.56$\pm$0.14       & 0.40$\pm$0.75   &  6.68                  & $3f_{10}-f_2+f_{\rm orb}$             \\ 
$f_{13}$     &  9.75701$\pm$0.00026   & 0.56$\pm$0.14       & 2.62$\pm$0.75   &  6.65                  & $3f_1-2f_5+2f_{\rm orb}$              \\ 
$f_{14}$     &  7.89790$\pm$0.00029   & 0.52$\pm$0.15       & 3.94$\pm$0.83   &  6.05                  & $f_5-f_{\rm orb}$                     \\ 
$f_{15}$     & 19.35892$\pm$0.00030   & 0.40$\pm$0.12       & 1.45$\pm$0.87   &  5.77                  &                                       \\ 
$f_{16}$     & 10.16518$\pm$0.00030   & 0.49$\pm$0.14       & 4.76$\pm$0.86   &  5.80                  & $f_5+2f_{\rm orb}$                    \\ 
$f_{17}$     & 14.09534$\pm$0.00023   & 0.44$\pm$0.10       & 1.38$\pm$0.67   &  7.43                  &                                       \\ 
$f_{18}$     & 15.46378$\pm$0.00029   & 0.44$\pm$0.13       & 3.94$\pm$0.83   &  6.04                  &                                       \\ 
$f_{19}$     & 13.87196$\pm$0.00023   & 0.42$\pm$0.09       & 3.35$\pm$0.66   &  7.57                  &                                       \\ 
$f_{20}$     &  3.02371$\pm$0.00019   & 0.44$\pm$0.09       & 3.39$\pm$0.56   &  8.94                  & $4f_{\rm orb}$                        \\ 
$f_{21}$     & 18.26412$\pm$0.00033   & 0.40$\pm$0.13       & 3.16$\pm$0.95   &  5.28                  & $7f_{10}-2f_2-f_5$                    \\ 
$f_{22}$     & 16.05471$\pm$0.00040   & 0.32$\pm$0.13       & 0.39$\pm$1.17   &  4.28                  &                                       \\ 
$f_{23}$     & 17.88336$\pm$0.00037   & 0.37$\pm$0.13       & 3.48$\pm$1.06   &  4.73                  &                                       \\ 
$f_{24}$     &  6.80183$\pm$0.00037   & 0.35$\pm$0.13       & 6.22$\pm$1.06   &  4.73                  & $9f_{\rm orb}$                        \\ 
$f_{25}$     & 20.11434$\pm$0.00032   & 0.34$\pm$0.11       & 5.61$\pm$0.92   &  5.48                  &                                       \\ 
$f_{26}$     &  7.43718$\pm$0.00042   & 0.32$\pm$0.13       & 1.67$\pm$1.23   &  4.09                  & $f_{10}-2f_{\rm orb}$                 \\ 
$f_{27}$     & 16.57356$\pm$0.00042   & 0.32$\pm$0.13       & 5.63$\pm$1.23   &  4.09                  &                                       \\ 
$f_{28}$     & 18.73168$\pm$0.00042   & 0.30$\pm$0.13       & 6.26$\pm$1.23   &  4.09                  & $2f_2-2f_1$                           \\ 
$f_{29}$     & 19.03705$\pm$0.00040   & 0.30$\pm$0.12       & 2.67$\pm$1.16   &  4.32                  &                                       \\ 
$f_{30}$     & 20.21511$\pm$0.00035   & 0.30$\pm$0.10       & 5.89$\pm$1.02   &  4.93                  &                                       \\ 
$f_{31}$     &  0.75212$\pm$0.00036   & 0.30$\pm$0.11       & 0.67$\pm$1.04   &  4.80                  & $f_{\rm orb}$                         \\ 
$f_{32}$     & 20.06306$\pm$0.00039   & 0.27$\pm$0.11       & 6.26$\pm$1.13   &  4.43                  & $2f_1+4f_{\rm orb}$                   \\ 
$f_{33}$     & 14.52484$\pm$0.00039   & 0.27$\pm$0.10       & 4.58$\pm$1.13   &  4.42                  & $f_2-f_5+7f_{\rm orb}$                \\ 
\hline
\multicolumn{6}{l}{$^a$ Frequencies are listed in order of detection. } \\
\multicolumn{6}{l}{$^b$ The noise was calculated in a range of 5 d$^{-1}$ around each frequency. } 
\end{tabular}
\end{table}

\bsp
\label{lastpage}
\end{document}